\documentclass[a4paper,11pt]{article}
\usepackage{jcappub} 

\newcommand{\nn}{\nonumber}
\newcommand{\corr}[1]{\langle #1 \rangle}

\title{\boldmath Curvature perturbations from vacuum transition during inflation}

\author[a]{Bj\"{o}rn Garbrecht,}

\author[b]{Pulkit S. Ghoderao}

\author[b]{and Arttu Rajantie}

\affiliation[a]{Physik Department, Technische Universit\"{a}t M\"{u}nchen, 85748 Garching, Germany}

\affiliation[b]{Department of Physics, Imperial College London, London SW7 2AZ, UK}

\emailAdd{garbrecht@tum.de, psg1618@imperial.ac.uk, a.rajantie@imperial.ac.uk}

\abstract{
We demonstrate that in the presence of a light scalar spectator field, vacuum transitions taking place during inflation can produce large, potentially detectable non-Gaussian signatures in the primordial curvature perturbation. Such transitions are common in theories with multiple scalar fields when the potential has several minima. Our computation proceeds by numerically finding the instanton solution that describes quantum tunnelling between vacuum states in a de Sitter background, calculating its dependence on the spectator field and, thereby, its effect on the expansion of space. For a scenario with Higgs inflation, we obtain the non-Gaussianity parameter $f_\text{NL} \sim O(10)$ and study its parameter dependence.
}

\begin{document}
\maketitle
\flushbottom

\section{Introduction}

In the original inflationary scenario \cite{Guth:1980zm}, inflation ends with a first-order phase transition. The inflaton field undergoes quantum tunnelling from a false vacuum to a true vacuum through the formation of bubbles. However, this scenario has the `graceful exit' problem whereby inflation never ends because the transition never completes. This problem essentially arises from the fact that the expansion rate of the false vacuum region separating true vacuum bubbles is higher than the expansion rate inside the bubbles, so that they can never meet and percolate. 

The widely accepted solution to the graceful exit problem is given by the new inflationary scenario \cite{Linde:1981mu,Albrecht:1982wi}, wherein a slowly-rolling scalar field is responsible for the exponential expansion as well as the generation of curvature perturbations in the early universe. The new inflationary scenario does not require there to be a first order phase transition, all the expansion can be generated by a potential that is flat enough. 

An alternative solution to the graceful exit problem pointed out in Ref.~\cite{Adams:1990ds} is to have a second field present during inflation that makes the vacuum decay rate field-dependent, so that percolation becomes favourable towards the end of inflation and the decay completes.  

In this work, we examine the possibility of a combination of both the above scenarios playing a part in the early universe. Namely that there is the standard slowly rolling inflationary field that generates the background exponential expansion and ends inflation when the slow-roll conditions fail, the evolution entering reheating era thereafter. In addition there is a second field that undergoes vacuum transition, but the energy released from the transition is not strong enough to end inflation. Thus we aim to answer the question: can a vacuum transition occurring during inflation generate significant curvature perturbations, in addition to the inflaton background, that can be observed through their non-Gaussianity?

The physical motivation for considering this question is the multiple vacua that arise in fundamental particle physics theories such as Grand Unified Theories (GUTs) \cite{ParticleDataGroup:2024cfk} and string theory \cite{Kachru:2003aw,Douglas:2006es,Baumann:2014nda}. It is possible that transition between these vacua takes place in the very early universe, during the inflationary stage, when fields take trans-Planckian values.   

One mechanism that generates significant non-Gaussian curvature perturbations during inflation is if there is a scalar field present in addition to the inflaton that is light in comparison to the inflationary Hubble rate. Looking at the inflationary evolution in terms of a separate universe picture \cite{Wands:2000dp}, the light field takes on different values in each separate Hubble volume during inflation. If the amount by which space expands depends on the local value of this field, these field fluctuations are converted into curvature perturbations. An example of this is the curvaton mechanism \cite{Moroi:2001ct, Lyth:2001nq, Lyth:2002my}, in which the spectator field comes to dominate the energy density towards the end of inflation.

Furthermore if the expansion of space is a non-linear function of the light field, which is generally the case when the field modulates some non-equilibrium process, these perturbations will also be significantly non-Gaussian. Examples of this that have been studied in the literature include the processes of preheating \cite{Finelli:1998bu,Bastero-Gil:2003usx, Suyama:2006rk,Chambers:2007se,Chambers:2008gu,Bond:2009xx,Chambers:2009ki,Imrith:2018uyk,Ghoderao:2023tow}, in which there is a non-linear non-perturbative transfer of energy from the inflaton to the spectator field after the end of inflation; oscillon formation \cite{Amin:2011hj}, in which non-linear solitons come to dominate the evolution post-inflation; vacuum transition, in which there is a non-perturbative change in the energy density during inflation due to a transition in the vacuum. This is the mechanism that we study in the present article. 

In the case of vacuum transition during inflation, the role of the light field can be played by the transition field~\cite{An:2024oui} or another scalar field. In this paper, we consider the latter scenario, which can be more natural in theories with multiple scalar fields because it does not require the transition field to be light.
     
Hence we illustrate the above mechanism through a generic model involving three fields: the \emph{inflaton} field generates the expansion, a vacuum \emph{transition} field alters the Hubble rate through a transition from false to true vacuum, and the value of a light \emph{spectator} field determines the decay rate of the false vacuum so that the transition happens at different times in separate Hubble volumes.

The multiple fields involved in our model of vacuum transition during inflation are again attributable to high energy particle physics models like string theory, which generally predict the existence of extra fields that can naturally play the role of the transition and the light spectator fields.

This article is divided as follows: After setting up generic assumptions on the nature of the potential, such as inflaton-dominated evolution and vacuum transition being favoured towards the end of inflation, we begin by developing the general formalism to calculate non-Gaussianity in Section~\ref{General formalism}. Working in the $\delta N$ formalism assuming a separate universe picture, we capture the non-Gaussianity through a sudden change in the inflaton's evolution due to vacuum transition.    

Including non-minimal coupling to gravity, in Section~\ref{Realistic model} we illustrate our formalism in the case of the Higgs inflation model. The decay rate is calculated numerically by implementing a shooting method assuming a fixed de Sitter background. Lastly, we delineate the parameter space and prove that decay never completes for a Hawking Moss transition for quartic potentials, which is the case for our model. Thus, we study the dependence of $f_\text{NL}$ on the mean value of the spectator field over our observable universe and its coupling to the transition field for a Coleman De Luccia transition.   

\section{Curvature perturbations from vacuum transition}
\label{sec:Curvature perturbations from vacuum transition}
\subsection{Overview}

Writing the spacetime metric as
\begin{align}\label{eqn:metric}
ds^2 = - dt^2 + e^{2\zeta(t,\vec{x})} a(t)^2 (dx^2 + dy^2 +dz^2),    
\end{align}
the curvature perturbation $\zeta$ captures local deviation of the scale factor from its background value $a(t)$. Thus it is a measure of the inhomogeneity in spatial slicings of the universe and as such can be directly related to the observed temperature anisotropy in the cosmic microwave background (CMB) radiation. From the metric \eqref{eqn:metric} the conformal time $\eta$ is defined as
\begin{align}
\eta = \int \frac{dt}{a(t)},
\end{align}
and the Hubble rate as 
\begin{align}
    H = \frac{1}{a(t)}\frac{da(t)}{dt}.
\end{align}

When pressure is a unique function of energy density, curvature perturbations arising from quantum fluctuations of the inflaton field are constant (frozen) on super-horizon scales~\cite{Lyth:2004gb}. However, if the system in consideration exhibits non-adiabatic pressure then curvature perturbations can evolve on super-horizon scales too \cite{Wands:2000dp}. A clear way to see this is by looking at the super-horizon evolution in real space through the separate universe picture.   

The separate universe approximation entails that the universe can be thought of as divided into Hubble volumes, each of which acts as a separate FRW universe. The fact that the universe can be divided into separate Hubble volumes is a straightforward consequence of causality, however that each Hubble volume remains FRW is the approximation.

In the separate universe picture, the curvature perturbation can be directly related to the local expansion in space through the $\delta N$ formalism \cite{Sugiyama:2012tj}. Evolving from a flat spatial slice of the universe with a scale factor $a_\text{ini}$ to a slice with uniform energy density $\rho_\text{ref}$, the $\delta N$ formula states that
\begin{align}\label{eqn:efolddefinition}
    \zeta = \delta N = N - \overline{N}, ~~\text{where}~ N = - \ln\left(\frac{a(\rho_\text{ref})}{a_\text{ini}}\right)
\end{align}
and $\overline{N}$ denotes the average of $N$ over the observable universe. Note that our definition for the function $N$ has an added negative sign as compared to prevalent definitions in the literature, for example in Ref.~\cite{Sugiyama:2012tj}, because $N$ denotes the number of efolds during inflation in our case.

The observable part of our universe corresponds to a scale that left the horizon $N_* \approx 55$ efolds before the end of inflation. Inflationary evolution on $N > N_*$ comprises of scales beyond the observable universe but which are still part of the entire universe. Between $N_*$ and the end of inflation $N=0$, there can be an intermediate time, lets call it $N_T$, at which the universe evolves non-adiabatically. In our case this would be when the vacuum transition occurs. To describe this, the observable universe can be divided into separate Hubble volumes in accordance with the separate universe picture.

For this process to produce curvature perturbations on large scales, there must be another light scalar field besides the inflaton, which develops significant fluctuations over the separate Hubble volumes. If we only had a single field, the inflaton, then it is possible to choose a cosmological gauge such that all fluctuations at scales much larger than $N_T$ can be made to disappear. Therefore we need a second scalar field in addition to the inflaton field, we refer to it as the spectator field $\chi$. Furthermore, this spectator field must also be light compared to the background Hubble rate $m_\chi < H$ so that its fluctuations are not damped on large scales.

These field fluctuations over the separate Hubble volumes must also modulate the local expansion of space. We need a non-adiabatic phenomenon to occur at $N_T$ so that each Hubble volume expands slightly differently and generates curvature perturbations in accordance with the $\delta N$ formula~\eqref{eqn:efolddefinition}. The non-adiabatic phenomenon we consider in the present article is vacuum transition.

In order to demonstrate the mechanism generating curvature perturbations from vacuum transition during inflation, the inflation, vacuum transition and spectator sectors are represented by three separate fields while constructing the general formalism in the next Section~\ref{General formalism}. However, it is important to note that this is just for illustration purposes. In reality we expect our mechanism to play a role in generating curvature perturbations for a wide range of beyond the Standard Model (BSM) scenarios that exhibit vacuum transition during inflation. Ultimately, by connecting such theoretically motivated BSM theories with the observations of non-Gaussianity in the CMB, we wish to test the suitability and delineate the parameter spaces of such theories to describe our universe. 

\subsection{Non-Gaussian curvature perturbations}
As remarked before, for illustration purposes we shall consider a scenario with three fields: the inflaton $\phi$, vacuum transition field $\sigma$ and spectator $\chi$. We shall assume that the inflaton dynamics dominates both the transition and spectator dynamics, and as such there is no back-reaction of the transition or spectator sectors on the inflationary background. 

If the transition field remains light throughout inflation until the transition point $N_T$, then we do not require a separate $\chi$ field to play the role of the spectator and generate significant fluctuations over the Hubble volumes in a separate universe picture. However, if the transition field is heavy then its fluctuations are suppressed on large scales and it does not contribute significantly to observable curvature perturbations. 

The $\delta N$ formalism \eqref{eqn:efolddefinition} in presence of spectator field $\chi$ yields a position dependent curvature perturbation \cite{Salopek:1990jq} 
\begin{align}\label{eqn:deltaNspectator}
    \zeta_\chi(\vec{x}) = \delta N(\chi(\vec{x})).
\end{align}
We assume that $\chi$ is a Gaussian random field and hence its statistics over the observable universe are fully characterised by the two point correlation function $\corr{\delta \chi(\vec{x}_1) \delta\chi(\vec{x}_2)}$, where $\delta \chi = \chi - \overline{\chi}$ and $\overline{\chi}$ is the mean value of the field over the observable universe. Assuming statistical isotropy and homogeneity, the two-point correlator in Fourier space becomes
\begin{align}
    \label{eq:Sigmadef}
    \corr{ \delta \chi(\vec{k}_1) \delta \chi(\vec{k}_2)} = (2\pi)^3\delta(\vec{k}_1+\vec{k}_2)\Sigma(k_1).
\end{align}
From which one can define the (reduced) power spectrum 
\begin{align}
    \mathcal{P}_\chi(k)  = \frac{k^3}{2\pi^2} \Sigma(k).
\end{align}
Then the probability distribution of the spectator field is a Gaussian
\begin{align}\label{eqn:probability distribution}
p_G(\chi)=\frac{1}{\sqrt{2\pi \corr{\delta\chi^2}}} \exp\left(-\frac{(\chi-\overline{\chi})^2}{2\corr{\delta\chi^2}}\right),
\end{align}
where the variance $\corr{\delta \chi^2}$ is given by
\begin{align}\label{eqn:Variance def}
    \corr{\delta \chi^2} = \int \frac{dk}{k}~\mathcal{P}_\chi(k) .
\end{align}

Even though the spectator field is Gaussian, it leads to non-Gaussian perturbations in the curvature through the $\delta N$ function in Eq.~\eqref{eqn:deltaNspectator} \cite{Lyth:2005fi}. The non-Gaussianity parameter for the curvature perturbations $f_\text{NL}$ is defined as~\cite{Komatsu:2001rj}
\begin{align}\label{eq:fNLdef}
	f_\text{NL}(\vec{k}_1,\vec{k}_2,\vec{k}_3) = - \frac{5}{6} \left(\frac{B_\zeta(\vec{k}_1,\vec{k}_2,\vec{k}_3)}{P_\zeta(\vec{k}_1) P_\zeta(\vec{k}_2) + P_\zeta(\vec{k}_1) P_\zeta(\vec{k}_3) + P_\zeta(\vec{k}_2) P_\zeta(\vec{k}_3)}\right),
\end{align}
where the curvature power spectrum $P_\zeta$ and bispectrum $B_\zeta$ in turn are defined by 
\begin{align}
\corr{\zeta(\vec{k}_1) \zeta(\vec{k}_2)} &= (2\pi)^3 \delta^{(3)}(\vec{k}_1+\vec{k}_2)P_\zeta(\vec{k}_1),\\
\corr{\zeta(\vec{k}_1)\zeta(\vec{k}_2)\zeta(\vec{k}_3)} &= (2\pi)^3 \delta^{(3)}(\vec{k}_1+\vec{k}_2+\vec{k}_3) B_\zeta(\vec{k}_1,\vec{k}_2,\vec{k}_3).
\end{align}
In typical slow-roll inflationary models, the inflaton contribution to the non-Gaussianity is proportional to the slow-roll parameters and consequently is small, $f_\text{NL} \lesssim 0.1$. However, the inflaton contribution to the power spectrum $P_\zeta$ is large and if the evolution is inflaton dominated, then we can approximate that all the contribution to the curvature power spectrum arises from the inflationary sector. The observed (reduced) curvature power spectrum at the comoving scale $k_*$ corresponding to the physical scale $k_\text{phys} = 0.05 \text{MPc}^{-1}$ \cite{Planck:2018nkj} is
\begin{align}
     \mathcal{P}_\zeta(k_*) = \frac{k_*^3}{2\pi^2}P_\zeta(k_*) = 2.1 \times 10^{-9} \equiv \mathcal{P}_*~.
\end{align}
By expanding the $\delta N$ formula Eq.~\eqref{eqn:deltaNspectator} in terms of the spectator field fluctuations 
\begin{align}
\label{eq:Taylor}
\delta N(\chi) = \delta N\Big|_{\chi = \overline{\chi}} + \delta N_{,\chi}\Big|_{\chi = \overline{\chi}} \delta\chi + \frac{1}{2} \delta N_{,\chi\chi}\Big|_{\chi = \overline{\chi}} \delta\chi^2,
\end{align}
its contribution to the bispectrum is obtained as \cite{Boubekeur:2005fj,Ghoderao:2023tow}
\begin{align}\nn
      	B_\zeta(\vec{k}_1,\vec{k}_2,\vec{k}_3)
    	&= \left(\delta N_{,\chi}\right)^2 \delta N_{,\chi\chi}\left(\Sigma(k_1)\Sigma(k_2)
    	+\Sigma(k_1)\Sigma(k_3)+\Sigma(k_2)\Sigma(k_3)\right)\\
        &~~~+ \left(\delta N_{,\chi\chi}\right)^3 \int \frac{d^3q}{(2\pi)^3} \Sigma(q)\Sigma(|\vec{q}+\vec{k}_1|)\Sigma(|\vec{q}+\vec{k}_1+\vec{k}_2|).
\end{align}
And hence to the non-Gaussianity $f_\text{NL}$ \eqref{eq:fNLdef} in the equilateral configuration $k_1=k_2=k_3 = k$ at the scale $k_*$ as \cite{Ghoderao:2023tow}
\begin{align}\label{fullfNL}
    f_\text{NL} = -\frac{5}{6} \frac{\mathcal{P}^2_\chi}{\mathcal{P}^2_*} (\delta N_{,\chi})^2 \delta N_{,\chi\chi} \Bigg|_{\chi = \overline{\chi}} -\frac{5}{6}  \frac{(\delta N_{,\chi\chi})^3}{\mathcal{P}^2_*}\int \frac{d\vec{q}}{4\pi}~\Sigma(q) \Sigma(|\vec{k}_* + \vec{q}|) \Sigma(|\vec{k}_*-\vec{q}|)  \Bigg|_{\chi = \overline{\chi}} ~.
\end{align}

\section{General formalism}
\label{General formalism}
\subsection{Setup}
Consider a potential $W(\phi,\sigma,\chi)$ comprised of three fields: the inflaton $\phi$, vacuum transition field $\sigma$ and a light spectator field $\chi$. On constant $\phi$ and $\chi$ slices, the potential in $\sigma$ admits two minima. The unstable minimum is known as the false vacuum and is denoted by $\sigma_{FV}(\phi,\chi)$. The stable minimum is the true vacuum denoted by $\sigma_{TV}(\phi,\chi)$. Henceforth we shall not indicate the dependence upon $\phi$ and $\chi$ explicitly and refer to the false and true vacuums simply as $\sigma_{FV}$ and $\sigma_{TV}$ respectively. The field $\sigma$ undergoes vacuum decay from its false vacuum to its true vacuum via a first-order phase transition that proceeds through the formation of true vacuum bubbles in the surrounding false vacuum region that eventually percolate and complete the transition.  

The decay rate per unit volume $\Gamma$ is given by
\begin{align}\label{eqn:decay rate definition}
    \Gamma = A e^{-B}\,
\end{align} 
where the exponential coefficient is given by the difference between Euclideanised action $S_E$ at a particular bounce solution $\sigma_b$ and in the false vacuum $\sigma_{FV}$, $B = S_E[\sigma_b] - S_E[\sigma_{FV}]$ \cite{Coleman:1977py}. On the other hand, the pre-factor $A$ is given at one-loop order by the determinant of fluctuations about the bounce~\cite{Callan:1977pt}. However, in practice one often takes $A \sim \mu^4_\text{eff}$ as a first approximation, where $\mu_\text{eff}$ captures the effective mass of the transition field.   

For our scenario, we assume initial conditions such that the transition field $\sigma$ is trapped in its false vacuum. The inflaton $\phi$ evolves in a slow-roll fashion, and the transition occurs at some point $\phi_T$ during its evolution. To simplify the treatment, we can restrict ourselves to only consider an inflaton-dominated approximation
\begin{align}\label{infdom}
    W(\sigma_{FV}) - W(\sigma_{TV}) \ll W(\sigma_{FV}).
\end{align}
In this case, the inflaton field will continue to evolve in a slow-roll fashion after the transition has occurred. This approximation also allows us to characterise the time evolution by the inflaton field value and to write the number of efolds in slow-roll in terms of the field evolution from $\phi_*$, which corresponds to the instance when the currently observed scale left the horizon to $\phi_\text{end}$, which corresponds to the end of inflation as \eqref{eqn:efolddefinition}
\begin{align}
    N(\chi)  = \int_{\phi_\text{end}}^{\phi_*} \frac{W}{W_{,\phi}} d\phi.
\end{align}   

\begin{figure}[t]
    \centering
    \includegraphics[scale=0.25]{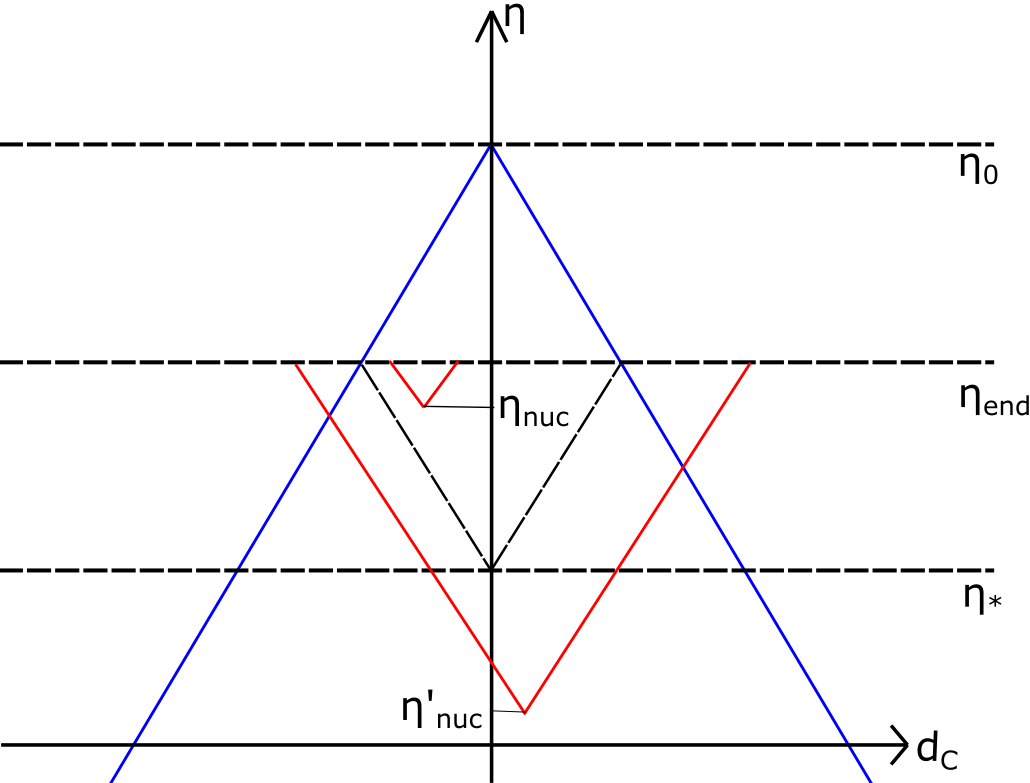}
    \caption{Spacetime diagram illustrating bubble nucleation in our past lightcone (in blue). The vertical axis denotes conformal time while the horizontal axis corresponds to comoving distance. $\eta_0$ is the conformal time today. For a bubble that nucleates at $\eta_\text{nuc}$ and expands at the speed of light, the forward light cone (in red) denotes the fraction of our observable universe covered by it at the end of inflation surface labelled by $\eta_\text{end}$. Similarly a bubble that nucleates at $\eta'_\text{nuc}$, before the largest currently observed scale left the horizon at $\eta_*$, covers the whole observable universe at $\eta_\text{end}$.}
    \label{fig:confbub}
\end{figure}

We consider the scenario in which the transition occurs during the last $N_*$ efolds, i.e. after the largest currently observable scale left the horizon, so that its effects are observable. See Figure~\ref{fig:confbub} for illustration. This means that no bubbles can have nucleated in our past lightcone prior to the instance given by $N_*$. The probability of this is given by $\exp(-\corr{n_\text{bub}})$, where $\corr{n_\text{bub}}$ is the expectation value of the number of bubbles nucleated in our past lightcone from the beginning of inflation $\eta_\text{begin}$ to the observed number of efolds $\eta_{*}$. It is given by \cite{Markkanen:2018pdo}
\begin{align}\label{nbub}
    \corr{n_\text{bub}} = \int d^4x~ \sqrt{-g} \Gamma = \int_{\eta_\text{begin}}^{\eta_{*}} d\eta~\frac{4\pi}{3} (\eta_0 - \eta)^3 a(\eta)^4 \Gamma(\eta),
\end{align}
where $\eta_0$ denotes conformal time today. Therefore we require 
\begin{align}\label{nobubcond}
    \corr{n_\text{bub}} \lesssim 1.
\end{align}

Using the definition of the number of efolds $dN = -Hdt = - H a d\eta$ and the scale factor during inflation $a(N) = a_\text{end} e^{-N}$, we find that $N$ e-foldings before the end of inflation, the conformal time has the value
\begin{align}
\eta(N) = \eta_\text{end} - \frac{1}{a_\text{end}} \int_{0}^{N} \frac{e^{N'}}{H(N')} dN',
\end{align}
where $\eta_{\rm end}$ is the conformal time at the end of inflation.
Substituting this back into the expression for the number of bubbles gives
\begin{align}
    \corr{n_\text{bub}} = \int_{N_*}^{N_\text{begin}} dN~\frac{4\pi}{3} \left(\eta_0 - \eta_\text{end} + \frac{1}{a_\text{end}} \int_{0}^{N} dN'~\frac{e^{N'}}{H(N')}\right)^3 \frac{a^3_\text{end}e^{-3N}}{H(N)} \Gamma(N).
\end{align}
For the standard $\Lambda$CDM universe, 
\begin{align}
    \eta_0 - \eta_\text{end} = \frac{1}{H_0} \int_0^{a_0} \frac{da}{a^2\sqrt{(\Omega_\Lambda + \Omega_m (a_0/a)^3 + \Omega_\text{rad} (a_0/a)^4) }} \approx 3.21 (a_0 H_0)^{-1},
\end{align}
where $\Omega_{\Lambda,m,{\rm rad}}$ refer to the present (at time $\eta_0$) ratios of energy density of vacuum energy, matter and radiation, respectively, with the total energy density in the universe. Then, in terms of the observed number of efolds
\begin{align}
    \frac{a_\text{end}H_\text{end}}{a_0 H_0} = e^{N_*}
\end{align}
the number of bubbles becomes
\begin{align}
    \corr{n_\text{bub}} = \int_{N_*}^{N_\text{begin}} dN~\frac{4\pi}{3} \left(3.21 + H_\text{end} \int_{0}^{N} dN'~\frac{e^{N'-N_*}}{H(N')} \right)^3 e^{-3(N-N_*)} \frac{\Gamma(N)}{H(N) H_\text{end}^3}.
\end{align}

We shall use the percolation parameter \cite{Guth:1980zm,Turner:1992tz,Copeland:1994vg}
\begin{align}\label{percolation parameter}
    p(N) = \frac{\Gamma(N)}{H(N)^4} 
\end{align}
to determine the time of the transition. The transition is taken to happen when the percolation parameter becomes unity and this is denoted by a subscript $T$ as $N_T$ or $\phi_T$, i.e.
\begin{align}
\label{completion:eof:inflation}
    p(N_T) = 1.
\end{align}
In this article, we only consider transitions that occur towards the end of inflation, $N_T \sim O(1)$, such that the transition is completed by bubbles whose sizes are several orders of magnitude smaller than the scale at which CMB measurements take place. Thus their effect on the CMB non-Gaussianity is only through the modulation of background inflationary expansion. Transitions that occur within the observable $N_*$ efolds but not towards the end of inflation can lead to interesting features in the post-inflationary evolution, but these are not the focus of the present article.

For simplicity we consider that the transition is completed predominantly by bubbles nucleated at $N_T$, i.e. the percolation parameter rises sharply at $N_T$,
\begin{align}\label{no back}
    \Big|\frac{dp}{dN}\Big|_{N=N_T} \gg 1.
\end{align}
Physically this translates to multiple bubbles forming inside one Hubble volume and coalescing within an inflationary time scale $H^{-1}$. In that case we can treat this transition as a sudden change in $\sigma$ from $\sigma_{FV}$ to $\sigma_{TV}$ at the level of the background inflationary evolution. With the sudden change in $\sigma$ at transition $\phi_T$, the number of efolds get split into
\begin{align}\label{dNintegral}
    \delta N = \int_{\phi_\text{end}}^{\phi_T}  d\phi  \frac{W}{W_{,\phi}} \Bigg|_{\sigma = \sigma_{TV}}  +\int_{\phi_T}^{\phi_*} d\phi  \frac{W}{W_{,\phi}} \Bigg|_{\sigma = \sigma_{FV}} .
\end{align}
By splitting the integral into two parts at $\phi_T$ we are also implicitly ignoring the background generated by bubbles nucleated at $\phi > \phi_T$, which is again justified under Eq.~\eqref{no back}.

\subsection{Non-Gaussianity} 
Our goal is to calculate the non-Gaussianity arising from the spectator field $\chi$. Assuming that the spectator field only negligibly contributes to the energy density during inflation, any direct dependence of the number of efolds \eqref{dNintegral} on $\chi$ through the potential can be ignored. Therefore it only indirectly depends on $\chi$ through $\phi_T(\chi)$, 
the time at which transition occurs in each Hubble volume,
\begin{align}\label{dNchiintegral}
    \delta N(\chi) = \int_{\phi_\text{end}}^{\phi_T(\chi)}  d\phi  \frac{W}{W_{,\phi}} \Bigg|_{\sigma = \sigma_{TV}}  +\int_{\phi_T(\chi)}^{\phi_*} d\phi  \frac{W}{W_{,\phi}} \Bigg|_{\sigma = \sigma_{FV}}.
\end{align}
Because the spectator field value is different in different Hubble volumes, each Hubble volume undergoes a different amount of expansion before the transition in it takes place. It is this difference in expansion between the Hubble volumes that can generate sizable non-Gaussianity.

To calculate $f_\text{NL}$ using Eq.~\eqref{fullfNL}, we need only the first and second derivatives of $\delta N (\chi)$ at $\chi = \bar{\chi}$.
But these can be immediately calculated without having to perform the integral~\eqref{dNchiintegral},
\begin{align}\label{dNderiv1}
\delta N_{,\chi} = \left( \frac{W}{W_{,\phi}} \Bigg|_{\sigma = \sigma_{TV}} - \frac{W}{W_{,\phi}} \Bigg|_{\sigma = \sigma_{FV}}  \right)_{\phi=\phi_T} \frac{d\phi_T}{d\chi} = \Delta \left(\frac{W}{W_{,\phi}}\right)\Bigg|_{\phi = \phi_T} \frac{d\phi_T}{d\chi},
\end{align}    
where we have introduced the notation $\Delta$ to indicate the difference between the true and false vacuum values. Similarly,
\begin{align}\label{dNderiv2}
\delta N_{,\chi\chi} = \Delta \left(\frac{\partial}{\partial \phi}\left(\frac{W}{W_{,\phi}}\right)\right)\Bigg|_{\phi = \phi_T} \left(\frac{d\phi_T}{d\chi}\right)^2 + \Delta \left(\frac{W}{W_{,\phi}}\right)\Bigg|_{\phi = \phi_T} \frac{d^2\phi_T}{d\chi^2}.
\end{align}

The remaining ingredients to fully determine the $\delta N$ derivatives are the derivatives of $\phi_T$ with respect to $\chi$, i.e. the change in time of transition across different Hubble volumes. These can be determined by numerically solving for the bounce solution with a different initial value for the spectator field corresponding to each separate Hubble volume. However, we can simplify the treatment without having to perform a separate bounce calculation for each Hubble volume for some specific types of potentials. For example, if we choose to capture the vacuum transition with a quartic potential and the coupling to the spectator field via a quadratic term,
\begin{align}
    U(\sigma,\chi) = \frac{\lambda'}{4!} \sigma^4 + \frac{g'}{3!} \sigma^3 + \frac{\mu^2}{2!} \sigma^2 + \frac{g^2_S}{2} \sigma^2 \chi^2.
\end{align}
Then we can define an ``effective" mass 
\begin{align}\label{mueffdef}
    \mu^2_\text{eff}(\phi,\chi) = \frac{\partial^2 U}{\partial \sigma^2}\Big|_{\sigma = \sigma_{FV}}
\end{align}
which is the only parameter in the quartic transition potential that changes as we move from one Hubble volume to the next. Note that the effective mass can also depend on the inflationary sector through a direct or indirect coupling between the inflaton and transition fields. For example, as we demonstrate in the next section for a Higgs inflation model, the inflaton and transition sectors can be coupled indirectly via a non-minimal coupling to gravity.

Now to calculate the derivatives of $\phi_T$ with respect to $\chi$, we know that $\phi_T(\chi)$ is a solution of the percolation condition Eq.~\eqref{completion:eof:inflation},
\begin{align}
\label{eq:per:cond}
    A({\mu_\text{eff}}) e^{-B({\mu_\text{eff}})} = H^4.
\end{align}
To proceed further without resort to numerical solutions, we ignore the $\phi_T$ dependence of $H$. Note that while $A$ and $B$ depend on all parameters in the potential for the transition field $\sigma$, here we are assuming that the dependence on the fields $\phi$ and $\chi$ enters only through the parameter $\mu_\text{eff}$. Then differentiating the percolation condition with respect to $\chi$ we have
\begin{align}
    \frac{d}{d\chi}{\mu_\text{eff}}(\phi_T,\chi) = 0
\end{align}
which implies
\begin{align}\label{eqn:dphiTdchi}
 \frac{d\phi_T}{d\chi} &= -\frac{{\mu_\text{eff}}_{,\chi}}{{\mu_\text{eff}}_{,\phi_T}},\\
\label{eqn:d2phiTdchi2}
 \frac{d^2\phi_T}{d\chi^2} &= -\frac{d}{d\chi}\left(\frac{{\mu_\text{eff}}_{,\chi}}{{\mu_\text{eff}}_{,\phi_T}}\right).   
\end{align}
We still need to obtain the transition point $\phi_T$ by finding out where the percolation parameter becomes unity i.e. by solving eq.~\eqref{completion:eof:inflation}. But we need to do so for only one value of $\chi$ at $\overline{\chi}$ and then use Eqs.~\eqref{eqn:dphiTdchi} and \eqref{eqn:d2phiTdchi2} above at that $\phi_T(\overline{\chi})$ and $\overline{\chi}$ as part of the non-Gaussianity calculation \eqref{fullfNL}.

\section{Higgs inflation model}
\label{Realistic model}

Here we demonstrate the use of our result~(\ref{fullfNL}) on an observationally viable scenario which is the extension of Higgs inflation~\cite{Bezrukov:2007ep} by a vacuum transition and a spectator field. The model is introduced in Subsection~\ref{subsec:Higgs:inf}, where we also review the dynamics of inflation before considering the vacuum transition. In Subsection~\ref{subsec:spectators}, the variance of the light spectator field due to vacuum fluctuations in quasi-de~Sitter space is calculated as well as its impact on $\delta N$ that enters the calculation for the non-Gaussianity. The theory of vacuum decay in de~Sitter space is reviewed in Subsection~\ref{subsec:vactrans} and applied to the present coupled model of Higgs inflation, which includes specifying the effective potential arising from the nonminimal coupling to gravity. Eventually, in Subsection~\ref{Parametric dependence}, we work out the dependence of $f_{\rm NL}$ on the model parameters. In the present phenomenological model, we find that the vacuum transition proceeds via subhorizon, Coleman--De~Luccia bubbles rather than Hawking--Moss transitions.

\subsection{Non-minimal coupling to gravity}
\label{subsec:Higgs:inf}

We consider the inflationary sector to be Higgs inflation \cite{Bezrukov:2007ep}, so the inflaton $\phi$ is the Higgs field. In Higgs inflation, $\phi$ is non-minimally coupled to gravity with the non-minimal coupling parameter $\xi_\phi$ being large $\xi_\phi \gg 1/6$. A non-minimal coupling to gravity must necessarily be included in the action due to QFT in curved spacetime considerations \cite{Tagirov:1972vv}. Therefore we can also include a non-minimal coupling for the transition and spectator fields through parameters $\xi_\sigma$ and $\xi_\chi$ respectively.

Curved spacetime QFT calculations show \cite{Markkanen:2013nwa} that the non-minimal coupling parameter possesses no fixed point under renormalisation group running, and therefore it has no preferred value theoretically. Its value also remains essentially unconstrained empirically~\cite{Atkins:2012yn}. For simplicity, we will assume that the non-minimal coupling for the transition field $\sigma$ has the same magnitude but opposite sign to the Higgs field, i.e., $\xi = \xi_\phi = -\xi_\sigma \gg 1/6$. Then a coupling between the Higgs field and the vacuum transition field arises naturally through the non-minimal term and no direct coupling between the fields is necessary. On the other hand, we assume that the spectator field $\chi$ is minimally coupled, i.e., $\xi_\chi=0$.
  
\begin{figure}[t]
    \centering
    \includegraphics[scale=0.33]{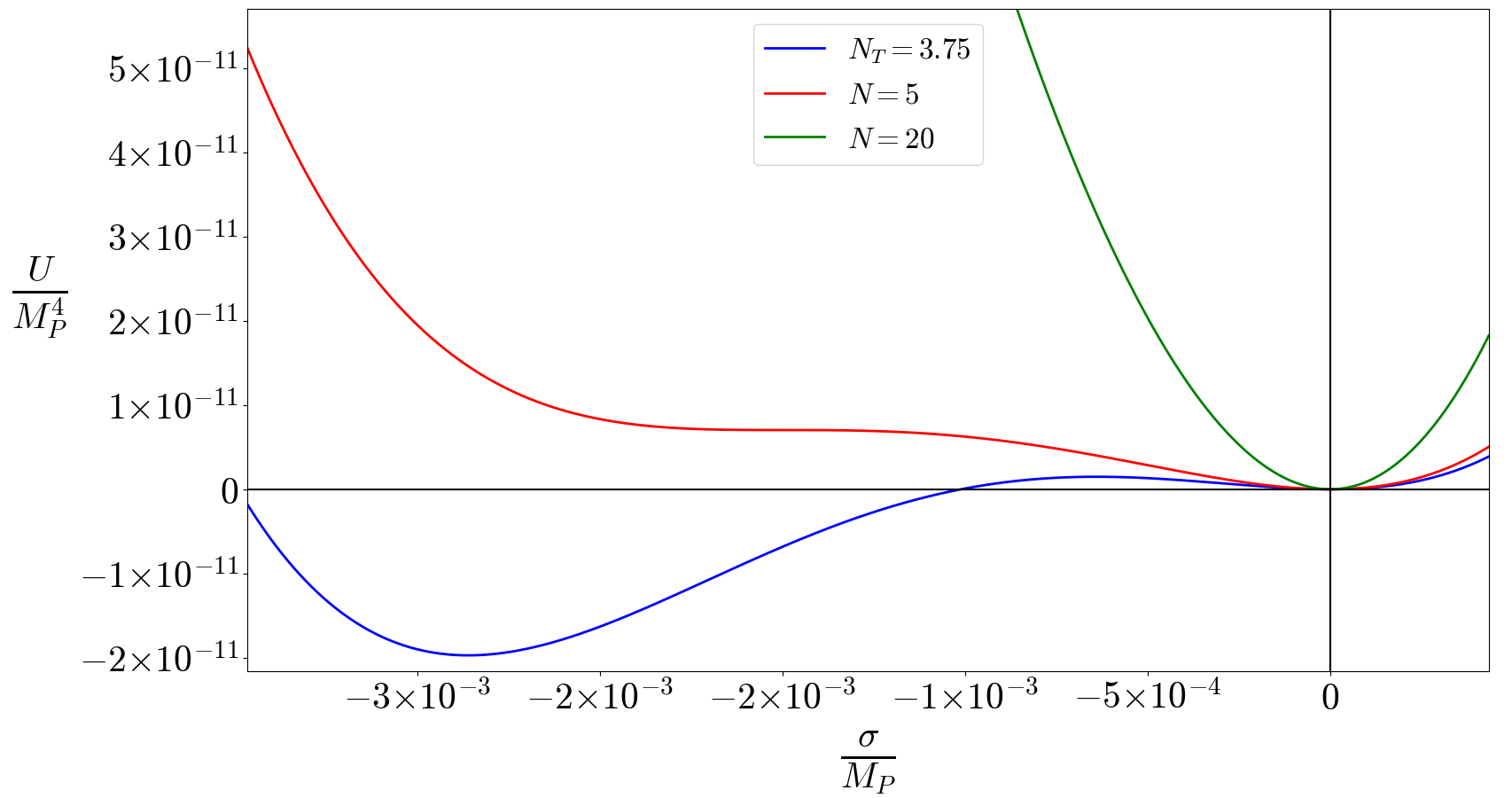}
    \caption{$U(\phi,\sigma,\chi=0)$ \eqref{eqn:Udef} at different times during inflation, labelled by the number of efolds $N$. This potential is chosen in such as way that when $\mathcal{R}$ is sufficiently high, it has a local minimum, a false vacuum, at $\sigma_{FV}=0$. The transition field parameters are $\lambda' = 100, g' =  10^{-1}M_P$ and $\mu = 5 \times 10^{-3}M_P$.}
    \label{fig:Uvsig}
\end{figure}

Let us therefore consider the action in Jordan frame
\begin{align}\label{eqn:Jordanframeaction}
&S = \int d^4x~\sqrt{-g}~\left(f(\phi,\sigma)\mathcal{R}- \frac{1}{2} g^{\mu\nu} \partial_\mu \phi \partial_\nu \phi - \frac{1}{2} g^{\mu\nu} \partial_\mu \sigma \partial_\nu \sigma - \frac{1}{2} g^{\mu\nu} \partial_\mu \chi \partial_\nu \chi  - W(\phi,\sigma,\chi) \right),
\end{align}
where
\begin{align}
f(\phi,\sigma) = \frac{M^2_P}{2} + \frac{1}{2} \xi \phi^2 - \frac{1}{2} \xi \sigma^2
\end{align}
and
\begin{align}
W(\phi,\sigma,\chi) = \frac{\lambda \phi^4}{4} + \lambda' \frac{\sigma^4}{4!} + g' \frac{\sigma^3}{3!} - \mu^2 \frac{\sigma^2}{2!} + \frac{1}{2} g^2_S \chi^2 \sigma^2.
\end{align}
Here $\mathcal{R}$ is the Ricci scalar. The transition field part of the potential is
\begin{align}\label{eqn:Udef}
    U(\phi, \sigma, \chi) = \lambda' \frac{\sigma^4}{4!} + g' \frac{\sigma^3}{3!} + \mu^2_\text{eff} \frac{\sigma^2}{2!},
\end{align}
where the effective mass of the transition field is given by
\begin{align}\label{eqn:mueff}
    \mu^2_\text{eff} = -\mu^2 + \xi \mathcal{R} + g^2_S \chi^2.
\end{align}
Its evolution during inflation is presented in Figure~\ref{fig:Uvsig}.  

Let us look at the action from the perspective of inflationary evolution of the background. For a majority of the evolution, the decaying field remains stuck in its false vacuum until the transition occurs at some value of $\phi_T$ towards the end of inflation. Therefore, we can make an approximation $\sigma \approx \sigma_{FV}$. Similarly, within one Hubble volume we assume that the spectator field $\chi$ is homogeneous and denote its value by $\chi_i$.
Therefore the kinetic terms for $\sigma$ and $\chi$ vanish. Then the action becomes
\begin{align}\label{eqn:back inflaton action}
    S \approx \int d^4x~\sqrt{-g}~\left(f(\phi,\sigma_{FV})\mathcal{R}
    + \frac{1}{2} \partial_\mu \phi \partial^\mu \phi- W(\phi,\sigma_{FV},\chi_i) \right).
\end{align}

The inflaton dynamics is best described in the Einstein frame where its slow roll equations of motion possess a simplified form. Making a conformal transformation of the metric
\begin{equation}\label{equ:conformaltfm}
    g_{\mu\nu} \to \tilde{g}_{\mu\nu} = \frac{2 f(\phi,\sigma_{FV})}{M^2_P} g_{\mu\nu},
\end{equation} the field develops a non-canonical kinetic term \cite{Kaiser:2010ps}
\begin{align}\nn
    S \approx \int d^4x~\sqrt{-\tilde{g}}~\Bigg(&\frac{M^2_P}{2}\tilde{\mathcal{R}} - \left(\frac{1}{1+\xi \frac{\phi^2}{M^2_P} - \xi \frac{\sigma^2_{FV}}{M^2_P}} + \frac{6 \xi^2 \frac{\phi^2}{M^2_P}}{(1 + \xi \frac{\phi^2}{M^2_P} - \xi \frac{\sigma^2_{FV}}{M^2_P})^2} \right) \frac{1}{2} \tilde{g}^{\mu\nu} \partial_\mu \phi \partial_\nu \phi\\
    &- \frac{1}{(1 + \xi \frac{\phi^2}{M^2_P} - \xi \frac{\sigma^2_{FV}}{M^2_P})^2} W(\phi,\sigma_{FV},\chi_i)  \Bigg).
\end{align}
We can convert the non-canonical kinetic term into canonical form by redefining the field $\phi \to \tilde{\phi}$ with
\begin{align}
    \frac{d\tilde{\phi}}{d\phi} = \sqrt{ \frac{1}{1  +\frac{\xi \phi^2}{M^2_P} -\xi \frac{\sigma^2_{FV}}{M^2_P}} + \frac{6 \xi^2 \phi^2}{\left(1 +\frac{\xi \phi^2}{M^2_P} - \xi \frac{\sigma^2_{FV}}{M^2_P} \right)^2}}.
\end{align}
This expression, along with the potential in the Einstein frame,
\begin{align}\label{eqn:fullEinsteinframePot}
    \tilde{W} \approx \frac{W(\phi,\sigma_{FV},\chi_i)}{(1 + \xi \frac{\phi^2}{M^2_P} - \xi \frac{\sigma^2_{FV}}{M^2_P})^2}
\end{align}
is sufficient to calculate the inflationary evolution numerically in the Einstein frame. 

Owing to the inflaton dominated assumption, we can neglect the false vacuum term $f(\phi,\sigma_{FV})\approx f(\phi) \equiv (M^2_P + \xi \phi^2)/2$ as a further simplification. 
Then the inflationary sector is identical to that of Higgs inflation \cite{Bezrukov:2007ep}. Therefore we fix the self coupling to its measured value for the Higgs field $\lambda \simeq 0.1$ \cite{ParticleDataGroup:2024cfk} and calculate the inflationary power spectrum and tensor to scalar ratio. Inflation ends when the slow-roll parameter becomes large,
\begin{align}
    \epsilon = \frac{M^2_P}{2} \left(\frac{\tilde{W}_{,\phi}}{\tilde{W}} \frac{d\phi}{d\tilde\phi}\right)^2 \sim 1,
\end{align}
which occurs at 
\begin{align}
    \phi_\text{end} = \sqrt{\frac{1}{2\xi}\left(\frac{\sqrt{192 \xi^2+32 \xi+1}}{6 \xi+1}-\frac{1}{6 \xi+1}\right)}.
\end{align}
The number of efolds are counted as
\begin{align}\nn
    N &= \int_{\phi_\text{end}}^{\phi} \frac{\tilde{W}}{\tilde{W}_{,\phi}} \left(\frac{d\tilde{\phi}}{d\phi}\right)^2 d\phi\\
    &= \left( \frac{\phi^2  (6 \xi+1)}{8} +\frac{1 - \sqrt{(8 \xi+1) (24 \xi+1)}}{16 \xi} -\frac{3}{4} \log \left(\frac{2 (6 \xi+1) \left(\phi^2 \xi+1\right)}{12 \xi+\sqrt{(8 \xi+1) (24 \xi+1)}+1}\right) \right)
\end{align}
from which we obtain $\phi_*$ for when the currently observed scale left the horizon $N_*=55$. Then the curvature power spectrum \cite{Lyth:2009zz}
\begin{align}
    \mathcal{P}_* = \frac{1}{24\pi^2} \frac{\tilde{W}}{\epsilon}\Big|_{\phi=\phi_*},
\end{align}
and tensor to scalar ratio \cite{Lyth:2009zz}
\begin{align}
    r = 16 \epsilon \Big|_{\phi = \phi_*}
\end{align}
can be determined. The observed inflationary power spectrum $\mathcal{P}_* = 2.1 \times 10^{-9}$ \cite{Planck:2018nkj} is obtained for $\xi = 15000$. For this value of the non-minimal coupling we have $\phi_\text{end} = 0.00877M_P$, $\phi_* = 0.0722 M_P$ and the tensor to scalar ratio $r = 0.0035$. This value lies well within the current observational bound $r < 0.03$ \cite{Tristram:2021tvh, Campeti:2022vom}. Thus our inflationary sector corresponds to Higgs inflation and is observationally viable.

\subsection{Spectator field fluctuations}
\label{subsec:spectators}
To calculate the non-Gaussianity, we need to examine the dynamics of the spectator field. Its variance over separate Hubble volumes is calculated in the Einstein frame. Keeping the field $\chi$ as dynamical, we have the action
\begin{align}\nn
    S \approx \int d^4x~\sqrt{-\tilde{g}}~\Bigg(\frac{M^2_P}{2}\tilde{\mathcal{R}} - \frac{1}{2} \tilde{g}^{\mu\nu} \partial_\mu \tilde{\phi} \partial_\nu \tilde{\phi} - \kappa(\tilde{\phi}) \frac{1}{2} \tilde{g}^{\mu\nu} \partial_\mu \chi \partial_\nu \chi
    - \tilde{W}(\tilde{\phi}, \chi)  \Bigg),
\end{align}
where the $\chi$ field now has a non-minimal kinetic term
\begin{align}
    \kappa(\tilde{\phi}) =  \left(1+\xi \frac{\phi(\tilde{\phi})^2}{M^2_P} - \xi \frac{\sigma^2_{FV}}{M^2_P}\right)^{-1},
\end{align}
and the potential in the Einstein frame is
\begin{align}
    \tilde{W}(\tilde{\phi},\chi) = \frac{W(\phi(\tilde{\phi}),\sigma_{FV}, \chi)}{(1 + \xi \frac{\phi(\tilde{\phi})^2}{M^2_P} - \xi \frac{\sigma^2_{FV}}{M^2_P})^2} .
\end{align}
Then the equations of motion for the fields are 
\begin{align}
    \ddot{\tilde{\phi}} + 3 \tilde{H} \dot{\tilde{\phi}} - \frac{\kappa_{,\tilde{\phi}}}{2} \dot{\chi}^2 + \tilde{W}_{,\tilde{\phi}} &= 0\\
    \ddot{\chi} + 3 \tilde{H} \dot{\chi} - \frac{\kappa_{,\tilde{\phi}}}{\kappa} \dot{\tilde{\phi}}\dot{\chi} + \frac{\tilde{W}_{,\chi}}{\kappa} &= 0.
\end{align}
Now assuming the background spectator field and its derivative are negligible during inflation, the super-horizon mode equation for the first-order fluctuation of the spectator is
\begin{align}
    \ddot{\delta\chi}_k + 3 \tilde{H} \dot{\delta\chi}_k + \frac{\kappa_{,\tilde{\phi}}}{\kappa} \dot{\tilde{\phi}} \dot{\chi}_k + \frac{1}{\kappa} \tilde{W}_{,\chi\chi} \delta\chi_k = 0.
\end{align}
We use the background slow-roll equations
\begin{align}
   3 \tilde{H} \dot{\tilde{\phi}} = -\tilde{W}_{,\tilde{\phi}}\quad\text{and}\quad \tilde{H}^2 = \frac{\tilde{W}}{3M^2_P} 
\end{align}
to obtain
\begin{align}
   \ddot{\delta\chi}_k + 3 \tilde{H} \left(1 - M^2_P\frac{\tilde{W}_{,\tilde{\phi}}}{3\tilde{W}}\frac{\kappa_{,\tilde{\phi}}}{\kappa}\right)\dot{\delta\chi}_k + \frac{1}{\kappa} \tilde{W}_{,\chi\chi} \delta\chi_k = 0.  
\end{align}

Before a mode exits the horizon, we take its evolution to be scale invariant and hence the power spectrum at horizon crossing as $\mathcal{P}(t_k) = \tilde{H}^2_k/4\pi^2$. Thereafter its evolution is described by the first order mode equation above and it develops an overdamped envelope $e^{-F(\phi_k,\phi_T)}$ from horizon crossing at $\phi_k$ to the transition at $\phi_T$. The overdamping coefficient is obtained as
\begin{align}\nn
    F(\phi_k,\phi_T) = \frac{3}{2} \int_{\phi_T}^{\phi_k}& d\phi~ \left(1 - M^2_P\frac{\tilde{W}_{,\phi}}{3\tilde{W}}\frac{\kappa_{,\phi}}{\kappa} \left(\frac{d\phi}{d\tilde{\phi}}\right)^2\right)\\
    & \times \left(1 - \sqrt{1 - \frac{4}{3} M^2_P \frac{\tilde{W}_{,\chi\chi}}{\kappa \tilde{W} }   \left(1 - M^2_P\frac{\tilde{W}_{,\phi}}{3\tilde{W}}\frac{\kappa_{,\phi}}{\kappa} \left(\frac{d\phi}{d\tilde{\phi}}\right)^2\right)^{-2}  }\right) \frac{\tilde{W}}{\tilde{W}_{,\phi}} \left(\frac{d\tilde{\phi}}{d\phi}\right)^2.
\end{align}
Hence the variance \eqref{eqn:Variance def} is given by
\begin{align}\label{varInt}
    \corr{\delta\chi^2} = \int \frac{dk}{k}~\mathcal{P}(k)= \int \frac{dk}{k} ~\frac{\tilde{H}^2_k}{4\pi^2} |e^{-F(\phi_k,\phi_T)}|^2 .
\end{align}

Using the slow-roll Friedmann equation and the horizon crossing relation $k = a_k \tilde{H}_k = a_\text{end} e^{-N_k} \tilde{H}_k$ we can rewrite the integral in terms of the Jordan frame field $\phi$ as
\begin{align}
    \corr{\delta \chi^2} = \int d\phi~ \left(\frac{\tilde{W}}{\tilde{W}_{,\phi}} \left(\frac{d\tilde{\phi}}{d\phi}\right)^2 - \frac{1}{2} \frac{\tilde{W}_{,\phi}}{\tilde{W}}\right)  \frac{\tilde{W}}{12\pi^2M^2_P} |e^{-F(\phi,\phi_T)}|^2 
\end{align}
where here and throughout we choose the scale factor to be unity at the end of inflation, $a_\text{end} = 1$. 

The variance at transition $\phi_T$ within our observable universe is calculated by choosing the appropriate integration limits
\begin{align}\label{obsVar}
   \corr{\delta \chi^2_T}  = \int_{\phi_T}^{\phi_*} d\phi~ \left(\frac{\tilde{W}}{\tilde{W}_{,\phi}} \left(\frac{d\tilde{\phi}}{d\phi}\right)^2 - \frac{1}{2} \frac{\tilde{W}_{,\phi}}{\tilde{W}}\right)  \frac{\tilde{W}}{12\pi^2M^2_P} |e^{-F(\phi,\phi_T)}|^2.
\end{align}
where $\phi_*$ corresponds to when the currently observed scale left the horizon. Similarly the cosmic variance, the variance of modes that lie outside our observable universe at transition (i.e. modes which left the horizon before $N_*$ efolds) is given by
\begin{align}\label{cosmicVar}
    \corr{\overline{\chi}^2}  = \int_{\phi_*}^{\phi_\text{begin}} d\phi~ \left(\frac{\tilde{W}}{\tilde{W}_{,\phi}} \left(\frac{d\tilde{\phi}}{d\phi}\right)^2 - \frac{1}{2} \frac{\tilde{W}_{,\phi}}{\tilde{W}}\right)  \frac{\tilde{W}}{12\pi^2M^2_P} |e^{-F(\phi,\phi_T)}|^2.
\end{align}

After populating the separate universes with a Gaussian distribution \eqref{eqn:probability distribution} with mean $\overline{\chi}$ and variance $\corr{\delta \chi_T^2}$ at transition, the non-Gaussianity parameter can be calculated using the $\delta N$-formalism \eqref{fullfNL}. However, the $\delta N$-calculation depends on time evolution and must take place in the Einstein frame. Therefore we need to include the full Einstein frame potential \eqref{eqn:fullEinsteinframePot} to calculate the derivatives of $\delta N$ \eqref{dNderiv1}--\eqref{dNderiv2} in the Einstein frame 
    \begin{align}
    \frac{\partial (\delta N)}{\partial \chi} &= \Delta \left(\frac{\tilde{W}}{\tilde{W}_{,\phi}} \left(\frac{d\tilde{\phi}}{d\phi}\right)^2 \right)\Bigg|_{\phi = \phi_T} \frac{d\phi_T}{d\chi} \\
    \frac{\partial^2 (\delta N)}{\partial \chi^2} &= \Delta \left(\frac{\partial}{\partial \phi}\left(\frac{\tilde{W}}{\tilde{W}_{,\phi}}\left(\frac{d\tilde{\phi}}{d\phi}\right)^2\right)\right)\Bigg|_{\phi = \phi_T} \left(\frac{d\phi_T}{d\chi}\right)^2 + \Delta \left(\frac{\tilde{W}}{\tilde{W}_{,\phi}} \left(\frac{d\tilde{\phi}}{d\phi}\right)^2 \right)\Bigg|_{\phi = \phi_T} \frac{d^2\phi_T}{d\chi^2}.
\end{align}

\subsection{Vacuum transition in de Sitter space}
\label{subsec:vactrans}
A nearly scale-invariant assumption for inflation implies the transition can be considered to occur in a fixed de Sitter background, $H \approx {\rm const}$. We find it is important to include the curved spacetime effect while calculating decay rate during inflation because the resulting bounce solution radius can be comparable to the Hubble radius. This effect has not been included in Ref.~\cite{An:2024oui} dealing with first order phase transitions during inflation, where the Minkowski decay rate is used.

The vacuum transition dynamics is best studied in the Jordan frame where the effective mass for the transition field is directly given by Eq.~\eqref{eqn:mueff}
where $\mathcal{R}$ can be obtained through the background action \eqref{eqn:back inflaton action} whose equations of motion for the inflaton field and metric are~\cite{Carroll:2004st}
\begin{align}
    \nabla_\alpha \nabla^\alpha \phi &= W_{,\phi} - \xi \phi \mathcal{R},\\
    \mathcal{R}_{\mu\nu} - \frac{1}{2} g_{\mu\nu} \mathcal{R} &= \frac{\partial_\mu \phi \partial_\nu \phi - \frac{1}{2} g_{\mu\nu} \partial_\alpha \phi \partial^\alpha \phi - g_{\mu\nu} W + (\nabla_\mu \nabla_\nu - g_{\mu\nu} \nabla_\alpha \nabla^\alpha) (M^2_P + \xi \phi^2 - \xi \sigma^2_{FV})}{M^2_P + \xi \phi^2 - \xi \sigma^2_{FV}}.
\end{align}
Contracting the metric equation with $g^{\mu\nu}$ and making use of the inflaton field equation we find
\begin{align}
    \mathcal{R} = \frac{ (1+6 \xi) \partial_\mu \phi \partial^\mu \phi + 4W + 6\xi \phi W_{,\phi}}{M^2_P + \xi (1 + 6\xi) \phi^2 - \xi \sigma^2_{FV}} .
\end{align}
Derivatives of the field can be ignored under the slow roll assumption, making 
\begin{align}
    \mathcal{R} \approx \frac{ (1 + 6 \xi)\lambda \phi^4 + 4 \left(\lambda' \frac{\sigma_{FV}^4}{4!} + g' \frac{\sigma_{FV}^3}{3!} - \mu^2 \frac{\sigma_{FV}^2}{2!} + g^2_S \chi^2_i\frac{\sigma_{FV}^2}{2}\right)}{M^2_P + \xi (1 + 6\xi) \phi^2 - \xi \sigma^2_{FV}}.
\end{align}
Owing to the high non-minimal coupling value $\xi \gg 1$ and the inflaton dominated approximation, this can be further simplified to
\begin{align}\label{eqn:approxR}
    \mathcal{R} \approx \frac{\lambda \phi^2}{\xi}.
\end{align}
Finally the effective mass \eqref{eqn:mueff} becomes
\begin{align}\label{realisticmueff}
    \mu_\text{eff}^2 \approx -\mu^2 + \lambda \phi^2 + g^2_S \chi^2,
\end{align}
which decreases towards the end of inflation as required for the vacuum transition to have observable effects. 

The fixed de Sitter background in our case has
\begin{align}
    H_{T} = \sqrt{\frac{\mathcal{R}(\phi_T)}{12}},
\end{align}
which implies $\phi$ can be fixed to its value at transition $\phi_T$ for the purposes of the decay rate calculation.

On a fixed de Sitter background, the transition field has an equation of motion \cite{Coleman:1980aw, Markkanen:2018pdo}
\begin{align}
    \frac{d^2\sigma}{dr^2} + 3 H\cot\left(H r\right)\frac{d\sigma}{dr} = \frac{dU}{d\sigma} 
\end{align}
in Euclideanised space with radius $r$ and with boundary conditions 
\begin{align}
\label{equ:bouncebc}
    \frac{d\sigma}{dr}\Big|_{r=0} = \frac{d\sigma}{dr}\Big|_{r= \pi/H} = 0~.
\end{align}

We calculate the decay rate \eqref{eqn:decay rate definition} numerically using Mathematica by employing a shooting algorithm on a domain $rH \in [10^{-3},\pi - 10^{-3}]$ and with $70$ digits of precision. Once the bounce solution $\sigma_b$ is obtained, its action is calculated by numerically integrating \cite{Cruz:2022ext}
\begin{align}\label{dSbounceaction}
    S_E[\sigma_b] = 2 \pi^2 \int_{0}^{\pi} dr~ \frac{\sin(r)^3}{H^2} \left( \frac{1}{2} \left(\frac{d\sigma}{dr}\right)^2 + \frac{U}{H^2}\right) .
\end{align}
The exponential coefficient in the decay rate $B$ is obtained by subtracting the constant part at false vacuum, $B = S_E[\sigma_b] - S_E[\sigma_{FV}]$. Unless otherwise specified, we always take the pre-factor in the decay rate to be $A = \mu_\text{eff}^4$ in numerical calculations.

\subsection{Parameter dependence of $f_\text{NL}$}
\label{Parametric dependence}
On a fixed de Sitter background, there exist two types of transitions: the Coleman--De~Luccia transition and the Hawking--Moss (HM) transition. However, depending on the value of Hubble rate, one dominates over the other. The Hawking--Moss transition only exists if \cite{Jensen:1983ac, Tanaka:1992zw}
\begin{align}\label{HM existence}
    H_{T} > \frac{\mu_\text{bar}}{2},
\end{align}
where $\mu_\text{bar} = \sqrt{|U''(\sigma_\text{bar})|}$. Here, $\sigma_\text{bar}$ is the value of the transition field at the top of the barrier separating true and false vacuums and prime denotes derivative with respect to the transition field.

We shall now show that the percolation parameter \eqref{percolation parameter} for a Hawking--Moss transition is always less than unity for a quartic potential. Denoting the HM decay rate as $\Gamma_\text{HM} = A_\text{HM}\exp(-B_\text{HM})$, the percolation parameter is bounded by
\begin{align}
    p = \frac{A_\text{HM}}{H_{T}^4}\exp(-B_\text{HM}) < \frac{A_\text{HM}}{H_{T}^4}
\end{align}
due to the exponent $B_\text{HM} > 1$. The prefactor for Hawking--Moss decay rate is given by~\cite{Camargo-Molina:2022paw}
\begin{align}\label{HMrate}
    A_\text{HM} = \frac{H^2_{T}}{6\pi} \mu_\text{eff} \mu_\text{bar},
\end{align}
so the bound on $p$ becomes
\begin{align}
    p < \frac{1}{6\pi} \frac{\mu_\text{eff} \mu_\text{bar}}{H^2_{T}} < \frac{1}{3\pi} \frac{\mu^2_\text{bar}}{H^2_{T}}, 
\end{align}
where to obtain the last inequality we use the fact that for a quartic transition potential, the maximum value of $\mu_\text{eff}$ can be $2 \mu_\text{bar}$ which occurs when $g'=0$. Now using the relation \eqref{HM existence} for the HM transition to exist we finally see that
\begin{align}
    p < \frac{4}{3\pi} < 1.
\end{align} 
Hence we conclude that a Hawking--Moss transition never completes during inflation (cf. Eq.~\eqref{completion:eof:inflation}) for an effectively quartic transition potential.   

Consequently, the parameter space of interest in our setup is reduced to 
\begin{align}\label{HMparameterrange}
     \mu_\text{eff} \gtrsim H_{T},
\end{align}
where the Coleman--De~Luccia transition dominates.

The effective mass for the Higgs inflation scenario is given by Eq.~\eqref{realisticmueff}. Because we are always assuming the dynamics of the $\chi$ field is subdominant to both the inflationary and decay sectors, the last term in this expression contributes negligibly to $\mu_\text{eff}$. The middle term, $\lambda \phi^2$, however gets fixed by the Higgs inflation parameter $\lambda \sim 10^{-1}$ and typical value of $\phi$ over the last $N_*$ efolds $\phi \sim 10^{-2}M_P$ giving a contribution to $\mu_\text{eff} \propto 10^{-3} M_P$. At higher values of $\phi$, the potential admits only one vacuum at $\sigma = 0$. To maintain the existence of the false vacuum towards the end of inflation when transition takes place, $\lambda \phi_T^2 - \mu^2 < (3/8) g'^2/\lambda'$. Therefore $\mu$ should lie close to $\lambda \phi_T^2$ and we choose $\mu = 5 \times 10^{-3}M_P$.

For an effective mass around $10^{-3}M_P$, the transition completes close to the end of inflation if the value of $g'$ is around $10^{-1} M_P$. But for such a high value of $g'$, in order to maintain the inflaton dominated simplification, one has to take $\lambda' \sim 100$. This value is not in the perturbative range and should be seen---just as $g'$---as a phenomenological parameter describing the tunnelling dynamics. One may, therefore, either assume that the latter occurs in a regime of non-perturbative couplings or construct a more sophisticated but perturbative sector describing the tunnelling field. One may also consider modifying the potential for $\phi$ beyond minimal Higgs inflation.

Following the above discussion, let us fix the parameter values for transition sector to be 
\begin{align}\label{eqn:transition field parameters}
    \lambda' = 100, g' = 10^{-1}M_P, \mu = 5 \times 10^{-3}M_P.
\end{align}

\begin{figure}[t]
    \centering
    \includegraphics[scale=0.3]{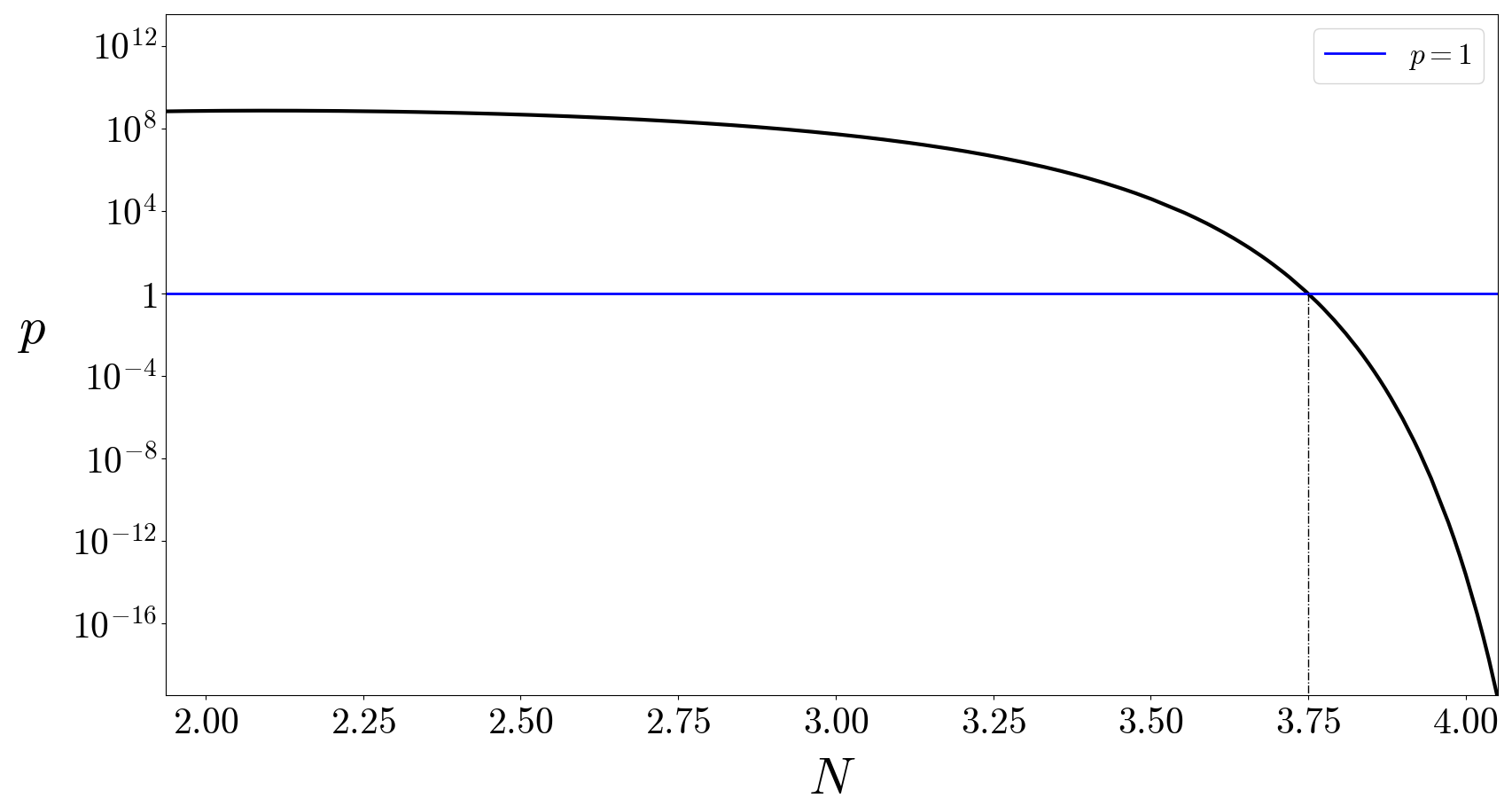}
    \caption{Percolation parameter during Higgs inflation as a function of the number of efolds and with transition field parameters $\lambda' = 100, g' =  10^{-1}M_P$ and $\mu = 5 \times 10^{-3}M_P$.}
    \label{fig:percparam}
\end{figure}

The plot of the percolation parameter as a function of number of efolds for these parameters is given by Figure~\ref{fig:percparam}.
We can see the transition completes at $N_T \approx 3.75$. We can also check that all the simplifications for our calculation to hold are satisfied. First, Eq.~\eqref{infdom} is satisfied, $(\tilde{W}(\sigma_{FV}) - \tilde{W}(\sigma_{TV})) / \tilde{H}^2_{T} M^2_P = 0.0093 \ll 1$, and therefore the evolution is inflaton dominated. Second, $\corr{n_\text{bub}} = 0$ as the potential admits only one vacuum above $N>5$ efolds, and therefore Eq.~(\ref{nbub}) is satisfied implying no bubbles can have nucleated before the largest currently observed scale left the horizon. Third, Eq.~(\ref{no back}) is satisfied, $|dp/dN|_{N=N_T} = 64.46 \gg 1$, and therefore the transition happens within one e-folding. 

Our mechanism can generate significant non-Gaussianity if the effective mass of the spectator field is light, i.e. $m_\chi = g_S \sigma_{FV} < H_T$. For the action \eqref{eqn:Jordanframeaction} with above transition field parameters, the effective mass of spectator field $m_\chi = g_S \sigma_{FV}$ is zero because $\sigma_{FV} = 0$. This makes the cosmic variance \eqref{cosmicVar} proportional to how long inflation lasted beyond $N > N_*$ efolds, and as such it depends on the specific inflationary scenario being considered. As the cosmic mean is drawn from a Gaussian distribution with variance equal to the cosmic variance, it further makes the cosmic mean an independent parameter. We shall only examine the cosmic mean over an arbitrary range $\overline{\chi} \in [0, 4\sqrt{\corr{\delta \chi^2_T}}]$. For the transition field parameters fixed above, the variance at transition \eqref{obsVar} is found to be
\begin{align}
    \corr{\delta \chi^2_T} = 4.5 \times 10^{-11} M^2_P.
\end{align}

The remaining parameter is the coupling between spectator and transition sectors $g_S$. We can check that over a range $g_S \in [0.1, 10]$ and $\overline{\chi} \in [0, 4\sqrt{\corr{\delta \chi^2_T}}]$ as before, the transition at $N_T = 3.75$ efolds is not affected. This is in keeping with the spectator dynamics being assumed sub-dominant to the transition dynamics. However, this allows us to obtain the variation in $f_\text{NL}$ as a function of the spectator sector parameters $g_S$ and $\overline{\chi}$ in the above range analytically directly using our non-Gaussianity calculation method expressed through Eq.~\eqref{fullfNL}. 

To simplify the calculation for the sub-leading term in $f_\text{NL}$ \eqref{fullfNL}, we divide the domain of integration into two: when $0 < q < k_*$, we approximate $\{\Sigma(|\vec{k}_* + \vec{q}|), \Sigma(|\vec{k}_*-\vec{q}|)\} \approx \Sigma(k_*)$ and when $k_* < q < q(N_T)$, we approximate $\{\Sigma(|\vec{k}_* + \vec{q}|), \Sigma(|\vec{k}_*-\vec{q}|)\} \approx \Sigma(q)$. Hence the total momentum integral is approximated as
\begin{align}
   \int \frac{d\vec{q}}{4\pi}~\Sigma(q) \Sigma(|\vec{k}_* + \vec{q}|) \Sigma(|\vec{k}_*-\vec{q}|) \approx \Sigma(k_*)^2 \int_{L^{-1}}^{k_*} q^2 dq~ \Sigma(q) + \int_{k_*}^{q(N_T)} q^2 dq~ \Sigma(q)^3,
\end{align}
and calculated numerically. Note that we have also introduced an infrared cut-off $L^{-1}$ corresponding to the size of the observable universe. In reality the cut-off, which arises from the $\delta N$ formalism, is not sharp. Therefore it is not possible to assign a precise value to $L$. But we know that the size of the observable universe as compared to the scale at which non-Gaussianity observations take place must satisfy $\ln(k_*L) \gtrsim 1$. The non-Gaussianity parameter $f_\text{NL}$ \eqref{fullfNL} increases with $\ln(k_*L)$, therefore to be on the conservative side we take $\ln(k_* L) = 1$ in our calculations to clearly demonstrate that non-Gaussianity owing to vacuum transition during inflation can be large.

\begin{figure}[t]
    \centering
    \includegraphics[scale=0.33]{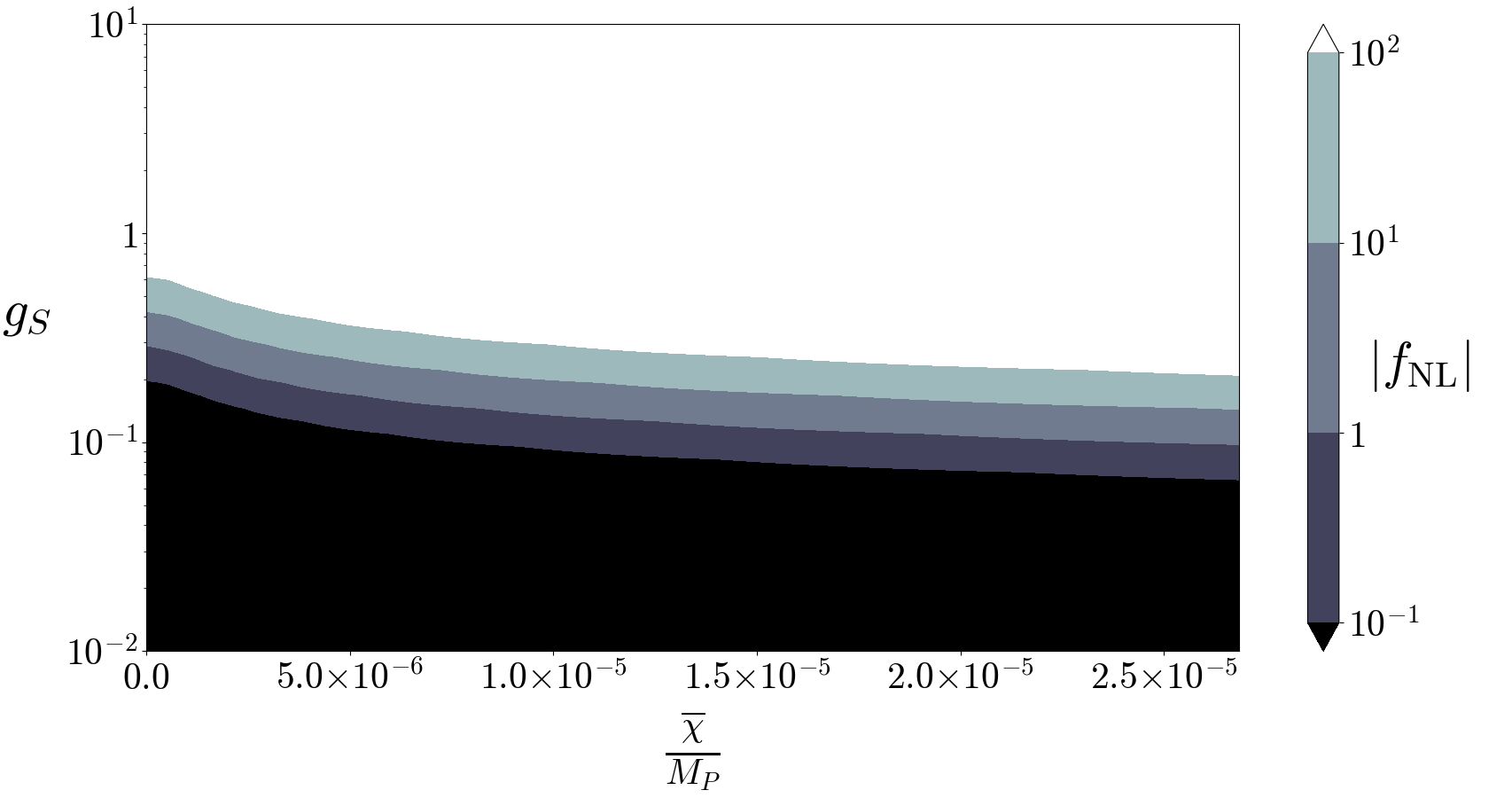}
    \caption{Non-Gaussianity parameter $f_\text{NL}$ as a function of  spectator coupling $g_S$ and cosmic mean $\overline{\chi}$. We highlight non-Gaussianity parameter values in the range $f_\text{NL} \in [0.1, 100]$ which might be accessible by current and future experiments. Other parameters in the Higgs inflation scenario are fixed to their sample values: $\lambda' = 100$, $g'= 10^{-1}M_P$ and $\mu = 5 \times 10^{-3}M_P$.}
    \label{fig:fNLvgSandchimean}
\end{figure}

In Figure~\ref{fig:fNLvgSandchimean}, we plot $f_\text{NL}$ as a function of the spectator sector parameters $g_S$ and $\overline{\chi}$. 
The dependence on $g_S$ reveals an approximately power law scaling which we find to be $f_\text{NL} \propto g^6_S$. On the other hand, the dependence on $\overline{\chi}$ is slight.

Finally let us summarise by providing a sample set of parameter values for the Higgs inflation scenario:
\begin{align}\label{sample parameters}
    \lambda' = 100, g' = 10^{-1}M_P, \mu = 5 \times 10^{-3}M_P, \overline{\chi} = 0, g_S = 0.5.
\end{align}
For these sample values the non-Gaussianity parameter is calculated to be
\begin{align}
    f_\text{NL} = -25.96~.
\end{align}

\section{Conclusion and Discussion}

We have outlined a general method to calculate the non-Gaussianity of primordial curvature perturbations from vacuum transition during inflation in the presence of a light scalar field. By using our method, one can obtain an analytical estimate of the non-Gaussianity parameter \eqref{fullfNL} for a generic potential through the difference in its derivatives before and after the transition \eqref{dNderiv1}--\eqref{dNderiv2}, and the derivatives of its effective mass for the transition field \eqref{eqn:dphiTdchi}--\eqref{eqn:d2phiTdchi2}.

It may be possible to relax some of the approximations used to obtain this analytic form, such as that the evolution is inflaton dominated, the decay occurs in the form of a sudden transition, there is no bubble background at transition, the number of efolds have no extrinsic dependence on the spectator field $\chi$ or that the Hubble rate at transition $H_{T}$ is independent of $\chi$. However, this would require us to extend the treatment and possibly perform the non-Gaussianity calculation numerically. And could be a direction for future work.   

By applying our method to the Higgs inflation model, we have conclusively shown that vacuum transition occurring during inflation in the presence of a light spectator field can be the source of large non-Gaussianity. We find that $f_\text{NL}$ takes values in the current observationally allowed range of $f^\text{local}_\text{NL} = -0.9 \pm 5.1$ \cite{Planck:2019kim} for a wide range of cosmic mean $\overline{\chi}$ values and a narrower range of spectator field couplings $g_S$. Higgs inflation model parameters which yield a $f_\text{NL}$ value larger than the observationally allowed range can already be ruled out. Cosmological experiments in the near future \cite{Heinrich:2023qaa} will further restrict this allowed range and can, therefore, more tightly constrain the parameter space of such models involving vacuum transition during inflation.

Finally, we note the requirement of a separate light spectator field in addition to the inflaton and vacuum transition fields in our model. For the transition field to be light, the top of the barrier must be flatter. This implies $|U''_\text{bar}| \approx |U''_{FV}|$ for a quartic decay potential, and hence the Hawking Moss transition, which is the dominant one for small effective mass, can never percolate as shown in Section~\ref{Parametric dependence}. Thus if the transition field is light then decay never completes and hence a separate light spectator field is required to generate the fluctuations that modulate the vacuum transition rate. 

However, if the decay potential contains operators higher than dimension four, then it could be possible to simultaneously satisfy the light effective mass and percolation conditions \cite{An:2024oui}. When the interactions and particle content required to have a vacuum transition during inflation scenario arise from integrating out sectors beyond the Standard Model, such higher dimensional operators may be generated. Finding well-motivated particle physics models where the transition field remains light can be a direction for further work.        

First order phase transitions occurring during inflation can also be the source for Gravitational Waves (GWs) \cite{Lopez:2013mqa,Ashoorioon:2015hya,Jiang:2015qor,Li:2020cjj,An:2022cce}. Along with the analysis presented here, a complementary examination of GWs could be decisive in identifying the non-Gaussian signal as the
result of a first order phase transition. Furthermore, it would help in constraining the allowed parameter range of theories exhibiting multiple vacua during inflation even better as upcoming GW experiments generate more data.

\appendix

\acknowledgments
PSG wishes to thank David Wands for useful discussion about the percolation parameter, the Physics Department at TUM and acknowledges support from the Imperial College Global Fellows Fund, Turing Scheme and the Government of Maharashtra, India. AR was supported by STFC Grant ST/X000575/1.

\bibliography{biblio}
\bibliographystyle{JHEP}
\end{document}